\documentclass[10pt,a4paper]{article}
\def\oggi{02.04.12}
\usepackage[english]{babel}
\usepackage[T1]{fontenc}
\usepackage[dvips]{graphicx}
\usepackage{amsmath, amsfonts, amsthm}
\usepackage{cite}
\usepackage{verbatim}

\numberwithin{equation}{section}

\newtheorem{hyp}{Assumption}
\newtheorem{teo}{Theorem}[section]
\newtheorem{defi}[teo]{Definition}
\newtheorem{proposition}[teo]{Proposition}
\newtheorem{lemma}[teo]{Lemma}
\newtheorem{coro}[teo]{Corollary}
\newtheorem{remark}[teo]{Remark}

\def\bx{\textbf{x}}
\def\by{\textbf{y}}
\def\be{\textbf{e}}

\def\bR{\textbf{X}}
\def\bu{\textbf{u}}
\def\bP{\textbf{P}}

\def\lag{\mathcal{L}}
\def\lagr{\mathcal{L}^{\ast}}
\def\Veff{\tilde{V}}
\def\C{\mathcal{C}}
\def\Sc{\mathcal{S}}
\def\M{\mathcal{M}}
\def\A{\mathcal{A}}
\def\F{\mathcal{F}}

\title{Asymptotic stability of synchronous orbits for a gravitating elastic sphere}
\author{Dario Bambusi, Emanuele Haus}

\begin{document}

\date{\oggi}
\maketitle

\begin{abstract}
We study the dynamics of an elastic body whose shape and position
evolve due to the gravitational forces exerted by a pointlike
planet. We work in the quadrupole approximation. We consider the
solution in which the center of mass of the body moves on a circular
orbit, and the body rotates in a synchronous way about its axis, so
that it always shows the same face to the planet as the Moon does with
the Earth. We prove that if any internal deformation of the body
dissipates some energy, then such an orbit is locally asymptotically
stable. The proof is based on the construction of a suitable system of
coordinates and on the use of LaSalle's principle. A large part of the
paper is devoted to the analysis of the kinematics of an elastic body
interacting with a gravitational field. We think this could have some
interest in itself.
\end{abstract}

\section{Introduction}

In this paper we study the dynamics of a deformable celestial body
interacting with a planet. In particular we are interested in
understanding the interaction of the internal degrees of freedom with
the orbital and spin ones.

The study of such a problem goes back to Darwin \cite{darwin1,darwin2}
(see also \cite{ferrazmello,efroimsky1}) and the theory he initiated
has been developed by many
authors\cite{alexander,goldreich,kaula,macdonald,peale}. Such a theory
consists of a procedure which allows to separate the internal degrees
of freedom (DOFs) from the orbital and spin ones. This is obtained by
showing that, in some approximation, the effect of the internal DOFs
is just that of producing an effective force acting on the orbital and
spin DOFs.  In particular one of the main issues of the theory is that
a dissipation acting on the internal DOFs induces an effective
dissipation on the orbital and spin DOFs (for a recent reference see
\cite{efroimsky2}). We emphasize that Darwin's procedure is heuristic,
and from a mathematical point of view, its range of validity is far
from being clear.

Our purpose in this paper is to prove the phenomenon of stabilization
of orbital and spin DOFs in a mathematically rigorous way, at least in
one simple model. Our model is the following one: we first approximate
the planet by a pure point. Then we model the satellite by an elastic
sphere, whose shape will change under the action of the gravitational
and dynamical forces.

From a dynamical system point of view, the system must be described by
a system of coupled differential equations governing both the
evolution of the orbital and spin DOFs and the internal (elastic) DOFs
of the satellite. In particular the equations of motion for the elastic
DOFs are of course partial differential equations.

The first problem we address is that of writing in a resonable way the
equations of motion of the system. It turns out that this is possible,
although nontrivial, by making very little assumptions, at least in the
quadrupole approximation. The introduction of suitable coordinates
occupies a large part of the paper.

Then we prove that, as expected, the so obtained equations have a
particular stationary solution in which the center of mass of the
satellite moves uniformly on a circular orbit and the satellite is
stretched in the direction of the planet (of course the shape of the
deformed body corresponds to the standard Love equilibria). Then we
prove that, if any internal deformation dissipates energy, then such a
stationary solution is asymptotically stable. In particular the
orbital DOFs relax to the circular ones. We emphasize
that our theory is local, so in particular it is valid only for small
initial values of the eccentricity.

We also would like to mention that our approach also applies to
satellites whose unperturbed shape is triaxial (like a rock), but in
order to conclude the proof one needs to substitute the argument of
section \ref{dissi} with a different argument which will be the object
of a separate paper.

We now briefly describe the proof and the structure of the paper. 

The starting point of the proof is the remark that, in quadrupole
approximation, the gravitational potential of an extended body in an
external gravitational field is a function only of the principal
moments of inertia and of the directions of the principal axes of
inertia. So it is natural to use such quantities as coordinates in the
configuration space of the elastic satellite. We first prove that it
is possible to complete such quantities to a system of
coordinates. Furthermore, due to the symmetries of the system, the
kinetic energy and the potential energy of the body have quite a
simple form.  (see eq. \eqref{eq:lagrangian}). For simplicity, we
restrict ourselves to the ``planar situation'' in which the center of
mass lies in a plane, the spin axis is orthogonal to such a plane and
the deformations are such that one of the principal axes of inertia of
the body is always orthogonal to the plane of the orbit. We first
study the dynamical system with such a Lagrangian proving the
existence of the above mentioned orbits, then we add dissipation and
use LaSalle's principle in order to get the main result. Concerning
dissipation, we only assume that any internal deformation of the body
produces some nonzero dissipation, and that the stress at a given time
is function of the strain and of its time derivative at that fixed
time, so that there are no memory effects.

We emphasize that the introduction of the coordinates is the most
difficult part of the proof, but we think that their construction
could have some interest in itself.

The main difficulties one has to face are of two kinds: the first one,
which will be discussed in detail in Section \ref{sect.coord}, is that
the separation between the rotational degrees of freedom and the
elastic ones relies on an arbitrary choice, which from a mathematical
point of view consists of the choice of a local section of a principal
fibre bundle. Concerning this point we emphasize that we are not
looking for a local separation, which could be obtained by the classical
decomposition of the displacement into strain tensor and local
skew-symmetric part of the displacement.  Instead we are looking for a
characterization of the deformations which globally do not rotate the
body. The difficulty is that there exist purely elastic deformations
with nonvanishing strain tensor which produce nontrivial rotations of
the body. The second difficulty is related to the fact that the
principal axes of inertia do not define directly a coordinate system,
since they are defined up to orientation. In order to overcome such
difficulty we use some properties of the spaces obtained as the
quotient of a Hausdorff space with respect to the action of a finite
group. The conclusion is that the wanted coordinates are a 24-fold
covering of the configuration space.

\vskip10pt

\noindent{\it Acknowledgements}. We thank Alessandra Celletti, Michael
Efroimsky, Sylvio Ferraz-Mello and Antonio Giorgilli for some
discussion on this problem, which led to considerable improvements of
the result.

\section{Coordinates in the configuration space}
\label{sect.coord}

In this section introduce suitable coordinates in the configuration space of the elastic body. Such coordinates allow to separate between elastic degrees of freedom and coordinates identifying the position and the orientation in space of the body. Though later, for simplicity, we will restrict ourselves to the planar situation, the construction of coordinates of the present section is fully general and is referred to three dimensional motion, with no planar constraints.

\subsection{General considerations}

We will use the Lagrangian description of elasticity. In this approach
one defines the so called material space $\Omega$, which is
essentially an abstract realization of the elastic body in some reference configuration.
In our case {\it we choose $\Omega$ to be a three dimensional sphere.}
In the following we will assume that the body is invariant under
rotations, in particular we assume that its density function
$\rho:\Omega\to\mathbb{R}^+$ is invariant under rotations. We denote
by $m$ the mass of the elastic body, i.e.
$$m=\int_{\Omega}{\rho(\bx)d^3\bx}\ .$$

The configuration of the body is a map $\zeta:\Omega \rightarrow
\mathbb{R}^3$, which gives the position in space of the the point
$\bx\in\Omega$.

\begin{remark}\label{identification}
Here and in the following, we identify the physical space, i.e. the
target of the configuration map $\zeta$, with the three-dimensional
real vector space $\mathbb{R}^3$.
\end{remark}

Of course the map $\zeta$ describes both the deformation
of the body and its position and rotation in space, so it
is natural to try to decompose $\zeta$ into a translation, a rotation and an internal deformation. To start, define 
the center of mass of the body by
\begin{equation} \label{eq:univoc1} \bR=\frac{1}{m}\int_{\Omega}
  {\zeta(\bx)\rho(\bx)d^3\bx }\ ,
\end{equation}
and decompose the configuration vector field $\zeta$ as
\begin{equation}
\zeta(\bx)=\bR+v(\bx)\ ,
\end{equation}
where $v$ is such that
\begin{equation} \label{c.m}
\int_{\Omega}{v(\bx)\rho(\bx)d^3\bx}=0\ .
\end{equation}
\begin{hyp}
As discussed in the introduction, we assume that $\bR$ lies in the plane generated by $\be_1$ and $\be_2$.
\end{hyp}
Here and below we denote by $\be_1$, $\be_2$, $\be_3$ the vectors of
the canonical base of $\mathbb{R}^3$.

Denote by $\mathcal{C}$ the space of the $v$'s such that (\ref{c.m})
holds. 

\begin{remark}
\label{C}
In principle, the space $\mathcal{C}$ should be an infinite dimensional function space, so
in order to discuss the dynamics one should introduce a suitable topology in
it, prove an existence and uniqueness theorem for the solutions of
the Cauchy problem and, in order to use energy conservation (or
dissipation) to prove dynamical properties, one should also prove that
dynamics is well posed in the energy space, which is in general
unknown \cite{marsdenhughes}.
\end{remark}

{\sl In the present paper we do not want to enter such a kind of mathematical
problems, so we cut the effective number of degrees of freedom to get an arbitrary but
finite number of variables, i.e. we make the following approximation.}

\textbf{Approximation.} \emph{We assume that the configuration space $\mathcal{C}$ has dimension $n+9$, i.e. we restrict the allowed deformations to a finite-dimensional space.}

\vskip10pt

Then we would like to factor out rotations in a way similar to
translations, however this requires a careful discussion. The
point is that it is clear what it means to rotate a body, but it is
not clear how to say that a deformation does not rotate the body: as
we will see, this is not a well defined concept.

To understand this point, we recall the standard analysis of the {\it
  local} deformation in elasticity theory. Fix a point
$\bx_0\in\Omega$. Then, the displacement vector field is defined by
$u(\bx):=\zeta(\bx)-\bx$. The gradient $\nabla u(\bx_0)$ is decomposed
as the sum of its symmetric part $\varepsilon$ and its skew-symmetric
part $\omega$. Then, \textit{at the point $\bx_0$}, the local
deformation is described by the strain tensor $\varepsilon$, while the
skew-symmetric part of the gradient $\omega$ describes the local
rotation. Thus, at a local level, the separation between deformation
and rotation is well-defined and completely standard.

On the other hand, when considering the global configuration, the situation is more complicated, because it is not trivial at all to answer the following question. \textit{Let the configuration $\zeta$ be assigned. Is the corresponding displacement vector field $u$ a ``pure deformation'', in the sense that it does not ``globally rotate the body'', or is it given by the composition of a rotation of the body with some ``pure deformation''?}

The answer to this question is easy if one considers only affine deformations, i.e. if one allows only displacements whose gradient is spatially constant. In fact, in this case, the rotation is simply described by the skew-symmetric part of the gradient of the displacement (evaluated at any point, since it is constant).

Instead we are here interested in general deformations which are
nonlinear. So the answer is not obvious as in the previous
situation. One could try to give some reasonable definitions of what a
``pure deformation'' is. For instance, we could fix $\bx_0\in\Omega$
and say that $u$ is a ``pure deformation'' if the displacement,
\textit{locally at $\bx_0$}, does not contain any rotation,
i.e. $\nabla u(\bx_0)$ is symmetric. Anyway, this seems quite
arbitrary; moreover, different choices of $\bx_0$ would result in
different definitions of ``pure deformation''. The is whether a
natural answer is possible or not.

We will conclude in a while that there is no natural way of defining
what it means that a displacement does not globally rotate the
body. On the contrary, it is trivial to explain what it means to
rotate a body. Mathematically, this corresponds to the fact that there
exists a group action $\A_1$ of the rotation group $SO(3)$ on the
configuration space $\C$, defined by \footnote{see Remark
  \ref{identification}}
\begin{eqnarray}
\nonumber \A_1: & SO(3) \times \mathcal{C} & \rightarrow \mathcal{C}\\ \label{g.a1}
& (\Gamma,v) & \mapsto \Gamma v\ ,
\end{eqnarray}
namely the configuration obtained by rotating the body in space with
the rotation $\Gamma$.  This group action allows one to introduce a
structure of principal fibre bundle in $\C$, the base manifold being
the quotient $\mathfrak{M}:=\C/SO(3)$. The elements of such a quotient
manifold are what one could call ``pure deformations''.

The fact that the quotient of a manifold under a group action is still a manifold is not always true, so we have to recall some basic facts about group actions in order to justify our assertion.

\begin{defi}
Let $G$ be a group acting on a set $X$. The action is said to be \textbf{free} if the following condition is verified: if there exist $x\in X$ and $g\in G$ such that $gx=x$, then $g$ is the identity.
\end{defi}

\begin{defi}
The action of a topological group $G$ on a topological space $X$ is said to be \textbf{proper} if the mapping $$G\times X\to X\times X \ (g,x)\mapsto (gx,x)$$ is proper, i.e., inverse images of compact sets are compact.
\end{defi}

The following classical results hold (see e.g. \cite{lee}).

\begin{teo}
Let $G$ be a topological group acting on a topological space $M$. If $G$ is compact, then the action is proper.
\end{teo}

\begin{teo}
If a Lie group G acts freely and properly on a smooth manifold, then the quotient $M/G$ is a smooth manifold.
\end{teo}

In our case, the fact that $SO(3)$ is a compact group and that the action $\A_1$ is obviously free implies that $\mathfrak{M}$ is a manifold.

As usual in this geometric context, it is useful to introduce coordinates in which a point of $\C$ is represented by an element of $SO(3)$ and an element of the base manifold. However a concrete representation of the elements of the quotient manifold can be obtained only locally, by introducing a local section of the bundle, namely by choosing a submanifold $\mathfrak{S}$ of $\C$, transversal to the group orbit. Now it is clear that there are infinitely many possible choices of the section of the bundle, which correspond to infinitely many admissible definitions of ``pure deformations''. Nonetheless, the physics is independent of the choice of the section $\mathfrak{S}$: a change in the choice of the section simply results in a change of coordinates. The fact that the section is only local is not a problem, as long as only small deformations are allowed.

Now, let $\A_1(SO(3))v_0$ be the orbit of the reference configuration under the action of the group $SO(3)$. Once the section $\mathfrak{S}$ has been chosen, it naturally induces a smooth one-to-one correspondence between a neighborhood of $\A_1(SO(3))v_0$ (obtained as the union of the orbits of the points of $\mathfrak{S}$ under the group action $\A_1$) and $SO(3)\times\mathfrak{S}$. Such a correspondence allows one to parameterize the space $\mathcal{C}$ of configurations through an element of the group $SO(3)$ and a point of the section $\mathfrak{S}$. The point is that, due to the isotropy of space, the Lagrangian of the body is going to be independent of the element of $SO(3)$ and will depend only on the point in $\mathfrak{S}$.

We remark that the machinery we have just introduced describes a general fact, which is independent of all the assumptions we will make later in the paper. We can summarize the result of our discussion in the following theorem:
\begin{teo}\label{th:coordinates.1}
Let $v_0\in\mathcal{C}$ be fixed and let $\mathfrak{S}$ be a section of $\mathcal{C}$ through $v_0$. Then there exist a neighborhood $\mathcal{U}\subset\mathcal{C}$ of $\A_1(SO(3))v_0$ and a one-to-one smooth function $f:SO(3)\times\mathfrak{S}\to\mathcal{U}$ with the property that $f(\Gamma,w)=\A_1(\Gamma)w$. Therefore, $\Gamma$ and $w$ can be used as coordinates on the configuration space $\mathcal{C}$.
\end{teo}

\subsection{Spherical symmetry and adapted coordinates}

In order to introduce explicitly the wanted set of coordinates, we recall the definition of principal moments and axes of inertia. The matrix of inertia of the body is a symmetric matrix
$I=\left\{I_{ij}\right\}_{i,j=1}^3$ whose elements are\footnote{see Remark \ref{identification}}
\begin{equation}
\label{in.1}
I_{ij}=I_{ij}(v):= \be_i\cdot \int_{\Omega} v(\bx) \wedge
(\be_j\wedge v(\bx)  ) \rho(\bx) d^3 \bx\ .
\end{equation}
{\sl The eigenvalues of such a matrix are called the principal moments
  of inertia and will be denoted by $I_1,I_2,I_3$. The eigenvectors of
  $I$ are called principal axes of inertia and will be denoted by
$\bu_1,\bu_2,\bu_3$.}

\vskip10pt

We make the assumption that the satellite is spherically symmetric, i.e.:

\begin{itemize}
\item[(i)] $\Omega$ is a three dimensional ball centered at the origin;
\item[(ii)] the corresponding density function $\rho_0(\bx)$ is a purely radial function of $\bx$, i.e. $\rho_0=\rho_0(\|\bx\|)$
\item[(iii)] the Lagrangian of the satellite (when ignoring the gravitational interaction with $M$) is invariant under the action $\mathcal{A}_1$ already introduced and also under the action $\mathcal{A}_2$ defined in the following way:
\begin{align}
\nonumber \A_2: SO(3) \times \mathcal{C} & \rightarrow \mathcal{C}\\ \label{g.a2}
(R,v) & \mapsto R v \circ R^{-1} \ .
\end{align}
\end{itemize}

\begin{remark}
This assumption implies that the satellite is spherical in its
reference configuration: in particular, all the principal moments of
inertia are equal. However, in order to obtain the stability of the
1:1 spin-orbit resonance, the fact that the principal moments of
inertia are not all equal plays a crucial role. As we will analyse
later, in our model the difference between the principal moments of
inertia is only a consequence of the elasticity of the satellite and
of the action of the tidal forces.
\end{remark}

The group action $\A_2$ has the following meaning. Imagine that the satellite is experiencing some deformation, which corresponds to a body configuration $v$. Then, applying $\A_2(R)$ to $v$ corresponds to producing a configuration which looks exactly like the previous one, except for the fact that the ``direction'' of the deformation inside the body has been rotated through the matrix $R$. We mean that if, for example, the
initial configuration is an ellipsoid with some principal axes, then the second one is an ellipsoid with the same shape, but with axes which have been rotated inside the body. This is a true elastic deformation that involves dissipation.

The action $\mathcal{A}_2$ is not free: in fact, consider for instance the reference configuration $v_0(\bx):=\bx$. It is immediate to verify that $\mathcal{A}_2(R)v_0=v_0$ for all $R\in SO(3)$. More in general, all body configurations which are symmetric with respect to a (continuous or discrete) subgroup of $SO(3)$ have a nontrivial stabilizer\footnote{the \textit{stabilizer} of $v\in\mathcal{C}$ under the group action $\mathcal{A}_2$ is the set of $R\in SO(3)$ s.t. $\mathcal{A}_2(R)v=v$} under the group action $\mathcal{A}_2$. Therefore, it is convenient to
consider also the action $\mathcal{A}_3$, combination of the actions
$\mathcal{A}_1$ and $\mathcal{A}_2$, defined by
\begin{equation}
\mathcal{A}_3(R)v:=\mathcal{A}_1(R)\mathcal{A}_2(R^{-1})v=v\circ R
\end{equation}
and study the couple of actions $\A_1$ and $\A_3$. It is easy to verify that the group action $\mathcal{A}_3$ is free.

We introduce now an adapted set of coordinates in a neighborhood of
the ``identical'' deformation $v_0$ (excluding however such a
configuration). To this end we need to introduce a few objects:

\begin{itemize}

\item[(1)] Define 
\begin{equation}
\mathcal{C}_{\neq}=\left\{v\in\mathcal{C}|I_1\neq
  I_2,I_1\neq I_3,I_2\neq I_3\right\}
\end{equation}  
and its complement
\begin{equation}
\mathcal{C}_{=}=\left\{v\in\mathcal{C}|I_i=I_j \ \text{for some}\ i\neq
j\right\}\ .
\end{equation}

This is useful since the principal axes $\bu_1$,
$\bu_2$, $\bu_3$ are uniquely determined in $\C_{\neq}$. 

\item[(2)] Define
\begin{equation}
\mathcal{D}:=\{v\in\C|I(v)\ \text{is diagonal}\}\ .
\end{equation}
We also define $\mathcal{D}_{\neq}:=\mathcal{D}\cap\C_{\neq}$.
Observe that $\mathcal{D}$ is a codimension 3 submanifold of $\C$,
invariant under the action $\A_3$ (we will show in the proof of Lemma
\ref{indep.action} that the action $\A_3$ leaves invariant the matrix
of inertia) and observe that $v_0\in\mathcal{D}$. Moreover, as we will
prove in Lemma \ref{indep.action}, on $\mathcal{D}\cap\C_{\neq}$ the
action $\A_1$ is independent of the action $\A_3$, and is
transversal to $\mathcal{D}$.

\item[(3)] Consider the group orbit $\A_3(SO(3))v_0\subset\mathcal{D}$,
  and let $\Sc\subset\mathcal{D}$ be a codimension 3 (in $\mathcal{D}$) manifold
  transversal to such a group orbit and passing through $v_0$. Actually we are interested in the
  restriction of such a section to a small neighborhood of $v_0$. We
  still denote by $\Sc$ such a {\it local} section. The existence of
  such an $\Sc$ is assured by the fact that the action $\A_3$ is free
  and therefore defines a foliation of $\mathcal{D}$.
  
\item[(4)] Finally define $\F$ to be the tube constituted by the orbits of
  $\A_1\circ\A_2$ starting in $\Sc\cap\C_{\neq}$, namely 
\begin{equation}
\label{effe}
\F:=\A_{1}(SO(3))\A_2(SO(3))(\Sc\cap \C_{\neq})\ .
\end{equation}
We remark that $\F=\A_{1}(SO(3))\A_3(SO(3))(\Sc\cap \C_{\neq})$.
\end{itemize}

\begin{remark} \label{I.sing}
The eigenvalues $I_j$, as functions of the matrix elements $\{I_{ij}\}$ (and therefore of the configuration $v$), are smooth functions on $\C_{\neq}$; however, the first derivatives of the $I_j$'s have a singularity at $\C_{=}$, therefore the $I_j$'s can be used as Lagrangian coordinates only on $\C_{\neq}$.
\end{remark}

\begin{remark}
When restricting to the submanifold $\mathcal{D}$, the eigenvalues $I_j$ coincide with the matrix elements on the main diagonal, and therefore they are obviously smooth functions of the configuration.
\end{remark}

We make the following assumption.

\begin{hyp} \label{i.indep}
We assume that the
functions $I_j:\mathcal{D}\to\mathbb{R}$, $j=1,2,3$ are independent in a neighborhood
of $v_0$.
\end{hyp}

\begin{remark}
The previous assumption is satisfied, for instance, if for any $j$ there exists a deformation which modifies $I_j$, leaving unaltered the other pincipal moments of inertis $I_k$ ($k\neq j$). The same assumption would not be satisfied if, for example, one added some additional constraint, like the incompressibility constraint. In that case, one would have to drop one degree of freedom.
\end{remark}

In the rest of the section we will prove the following Theorem:
\begin{teo}
\label{coordinates}
There exist functions $(z_1,z_2,\ldots,z_n)$, with
$$z_j:\Sc\rightarrow\mathbb{R} \quad (j=1,2,\ldots,n)$$
such that:
\begin{itemize}
\item[(i)] $(I_1,I_2,I_3,z_1,z_2,\ldots,z_n)$ is a set of
  smooth coordinates on $\Sc$.
\item[(ii)] the map
\begin{equation}\label{represent}
 SO(3)\times SO(3)\times(\Sc\cap\C_{\neq})\ni (\Gamma,R,I_1,I_2,I_3,z_1,z_2,\ldots,z_n)\mapsto \A_1(\Gamma)\A_2(R)w\in\F
\end{equation}
is a 24-fold covering of $\mathcal{F}$; here we denoted
$w=(I_1,I_2,I_3,z_1,z_2,\ldots,z_n)$.
\end{itemize}
\end{teo}

\begin{remark}
\label{r.24}
The number 24 arises as the order of the chiral octahedral group. More
precisely it corresponds to the number of ways in which an oriented
triple of orthonormal vectors can be rotated in such a way that the
vectors lie on a triple of unoriented fixed orthogonal axes.
\end{remark}

\subsubsection{Proof of Theorem \ref{coordinates}}

To begin with, we prove the existence of $(z_1,z_2,\ldots,z_n)$ satisfying
(i). Observe that $\Sc$ is a smooth submanifold of $\C$. Then, since
$I_1(w),I_2(w),I_3(w)$ are independent functions, it is possible to
complete the triple $(I_1,I_2,I_3)$ to a local system of coordinates
$(I_1,I_2,I_3,z_1,z_2,\ldots,z_n)\in\Sc$ near $v_0$. 

In the rest of the section, we will prove (ii). As a first step, we
show that the two actions of $SO(3)$ on $\mathcal{C}_{\neq}$ are
independent, which is implied by the following Lemma.

\begin{lemma} \label{indep.action}
For any fixed $\hat v\in\C_{\neq}$, we consider two subspaces of $T_{\hat v}\C_{\neq}$, tangent to the group orbits $\A_1(SO(3))\hat v$ and $\A_3(SO(3))\hat v$, namely $$\mathcal{T}_1:=T_{\hat v}\A_1(SO(3))\hat v\qquad\mathcal{T}_3:=T_{\hat v}\A_3(SO(3))\hat v\ .$$ Then, $\mathcal{T}_1\cap\mathcal{T}_3=\{0\}$. Moreover, if $\hat{v}\in\C_{\neq}\cap\mathcal{D}$, then $\mathcal{T}_1$ is transversal to $\mathcal{D}$.
\end{lemma}
\begin{proof}
In order to prove the thesis, we start by showing that the action $\A_1$ rotates the matrix of inertia, while the action $\A_3$ leaves it invariant.
We have
\begin{eqnarray}
\nonumber I_{ij}(\Gamma\hat{v}) & = & \be_i \cdot \int_{\Omega}{\Gamma\hat{v}(\bx)\wedge(\be_j\wedge \Gamma\hat{v}(\bx))\rho(\bx)d^3\bx}\\
& = & \Gamma^{-1}\be_i \cdot \int_{\Omega}{\hat{v}(\bx)\wedge(\Gamma^{-1}\be_j\wedge\hat{v}(\bx))\rho(\bx)d^3\bx}\ ,
\end{eqnarray}
which shows that the $I(\A_1(\Gamma)\hat{v})$ is the matrix of $I(\hat{v})$, just referred to a rotated basis.
On the other hand, we have
\begin{equation}
I_{ij}(\hat{v}\circ R)=\be_i \cdot \int_{\Omega}{\hat{v}(R\bx)\wedge(\be_j\wedge \hat{v}(R\bx))\rho(\bx)d^3\bx}\ .
\end{equation}
Setting $\by=R\bx$, we have
\begin{equation}
I_{ij}(\hat{v}\circ R)=\be_i \cdot \int_{\Omega}{\hat{v}(\by)\wedge(\be_j\wedge \hat{v}(\by))\rho(\by)d^3\by}\ ,
\end{equation}
which means that the action of $\A_3$ leaves the matrix of inertia invariant.
This implies
\begin{equation}
d I(\hat{v})v_3=0\ \forall v_3\in\mathcal{T}_3\ ,
\end{equation}
while
\begin{equation}
d I(\hat{v})v_1 \neq 0\ \forall v_1\in\mathcal{T}_1\setminus\{0\}\ , 
\end{equation}
from which the independence follows.

To get the transversality when $\hat{v}\in\C_{\neq}\cap\mathcal{D}$, we remark that $\A_1$ rotates the principal axes of inertia, then, since the three eigenvalues are distinct, it destroys the diagonal structure of $I$.
\end{proof}

\begin{remark}
As an obvious corollary of Lemma \ref{indep.action}, we also have that $\A_1$ and $\A_2$ are independent at any point $\hat{v}\in\C_{\neq}$, in the sense that $\mathcal{T}_1$ is transversal to the tangent space $\mathcal{T}_2:=T_{\hat v}\A_2(SO(3))\hat v$.
\end{remark}

Moreover, we observe that in $\C_{\neq}$ the three eigenvalues of the matrix of inertia are distinct, so the eigenvectors $\bu_1$, $\bu_2$, $\bu_3$ are well determined. Furthermore the dependence of the eigenvalues and eigenvectors on the configuration $v\in\C_{\neq}$ is smooth.

Now, we want to represent any configuration $v\in\F$ in the form 
\begin{equation} 
\label{decomposition}
v=\mathcal{A}_1(\Gamma) \mathcal{A}_2(R) w\ ,\quad
w\in\Sc\cap\C_{\neq}\ .
\end{equation}
Let us first represent any
$v\in\C_{\neq}$ in the form
\begin{equation} \label{decomposition2}
v=\A_1(\tilde{\Gamma})\tilde{w}\ ,\quad
\tilde{w}\in\mathcal{D}_{\neq}\ .
\end{equation}

\begin{proposition}
\label{covering}
The map
\begin{eqnarray*}
\Pi:  SO(3) \times\mathcal{D}_{\neq} & \rightarrow &
\C_{\neq}\\  (\tilde{\Gamma},\tilde{w}) & \mapsto &
\mathcal{A}_1(\tilde{\Gamma}) \tilde{w}
\end{eqnarray*}
is a 24-fold covering map.
\end{proposition}

The proof will make use of the following Theorem, which
is an immediate corollary of \cite{hatcher}, Proposition 1.40, p. 72.
\begin{teo} \label{th:cover}
If $G$ is a finite group, acting freely on a Hausdorff space $X$, then
the quotient map $X\rightarrow X/G$ is a covering map.
\end{teo}

\begin{proof}[Proof of Proposition \ref{covering}]
We observe that each $v\in\C_{\neq}$ has many distinct
representations of the form (\ref{decomposition2}): since the three
principal moments of inertia are distinct from one another, the
directions of the principal axes of inertia are well-determined, but the
same is not true for what concerns their orientation; moreover, any of the principal axes may
be labeled $\bu_1$ as well as $\bu_2$ or $\bu_3$.
In order to make this rigorous, consider the equation
\begin{equation}
\mathcal{A}_1(\tilde{\Gamma}_1)\tilde{w}_1=\mathcal{A}_1(\tilde{\Gamma}_2)\tilde{w}_2\ ,
\end{equation}
with $\tilde{w}_1,\tilde{w}_2\in\mathcal{D}_{\neq}$. This implies
\begin{equation}
\tilde{w}_2=\mathcal{A}_1(\tilde{\Gamma}_2^{-1}\tilde{\Gamma}_1)\tilde{w}_1\ .
\end{equation}
Therefore, since $\tilde{w}_1,\tilde{w}_2\in\mathcal{D}_{\neq}$, the rotation $\tilde{\Gamma}_2^{-1}\tilde{\Gamma}_1$ must transform the set $\{\bu_1,\bu_2,\bu_3\}$ to a set of unit vectors having the same directions. It is easy to see that the set of rotations satisfying this property is the subgroup of $SO(3)$ generated by the three rotations
$$
R_1=\left[ \begin{array}{ccc}
1 & 0 & 0 \\
0 & 0 & -1 \\
0 & 1 & 0 \end{array} \right]
$$
$$
R_2=\left[ \begin{array}{ccc}
0 & 0 & 1 \\
0 & 1 & 0 \\
-1 & 0 & 0 \end{array} \right]
$$
$$
R_3=\left[ \begin{array}{ccc}
0 & -1 & 0 \\
1 & 0 & 0 \\
0 & 0 & 1 \end{array} \right]\ .
$$
Such a subgroup, which we will denote by $O$, is isomorphic to the group of the orientation preserving symmetries of the cube, which is a group of order 24, known as the \emph{chiral octahedral group}. This argument shows that the possible representations of the form (\ref{decomposition2}) are at most 24. On the other hand, for any $\tilde{w}\in\mathcal{D}_{\neq}$ and $\tilde{\Gamma}\in SO(3)$, we have that the expression
\begin{equation}
\mathcal{A}_1(\tilde{\Gamma}\Gamma_O)[\mathcal{A}_1(\Gamma_O^{-1})\tilde{w}]
\end{equation}
yields 24 different representations of the same configuration, as $\Gamma_O$ varies within the group $O$. Therefore, each configuration $v\in\C_{\neq}$ has exactly 24 distinct representations of the form (\ref{decomposition2}) and a natural identification arises between $\mathcal{C}_{\neq}$ and $(SO(3)\times\mathcal{D}_{\neq}/O$, where the action of $O$ on $SO(3)\times\mathcal{D}_{\neq}$ is defined by
\begin{equation}
[\Gamma_O,(\tilde{\Gamma},\tilde{w})]\mapsto(\tilde{\Gamma}\Gamma_O,\mathcal{A}_1(\Gamma_O^{-1})\tilde{w})\ .
\end{equation}
Now, applying Theorem \ref{th:cover} with $X=SO(3)\times\mathcal{D}_{\neq}$ and $G=O$, we get the thesis.
\end{proof}

The above Proposition given as a global statement applies also to a small tube of orbits originating in $\Sc$. Precisely, define $\mathcal{T}:=\A_3(SO(3))(\Sc\cap\C_{\neq})$: then we have
\begin{coro}
\begin{eqnarray*}
\Pi: SO(3)\times\mathcal{T} & \rightarrow &
\F\\(\tilde{\Gamma},\tilde{w}) & \mapsto &
\mathcal{A}_1(\tilde{\Gamma}) \tilde{w}
\end{eqnarray*}
is a 24-fold covering map.
\end{coro}
\begin{proof}
The only thing we have to prove is that $\F$ is the image of $SO(3)\times\mathcal{T}$ through $\Pi$. However, this is obvious, since 
\begin{equation}
\Pi(SO(3)\times\mathcal{T})=\A_1(SO(3))(\mathcal{T})=\A_1(SO(3))\A_3(SO(3))(\Sc\cap\C_{\neq})=\F\ .
\end{equation}
\end{proof}

\noindent
{\it End of proof of Theorem \ref{coordinates}.} The last step consists in factoring out the group action $\A_3$. This is easy, since the action $\A_3$ is free. Therefore, one can decompose
\begin{equation}
\mathcal{T}\ni\tilde{w}=\A_3(\tilde{R})w \quad (w\in\Sc\cap\C_{\neq})
\end{equation}
in a unique way, and moreover the map
$$\tilde{w}\mapsto(\tilde{R},w)$$
is smooth. Therefore, a 24-fold covering of $\F$ is naturally induced by the map
\begin{equation}
(\tilde{\Gamma},\tilde{R},w)\mapsto \A_1(\tilde{\Gamma})\A_3(\tilde{R})w\ .
\end{equation}
Now, setting
$$\Gamma:=\tilde{\Gamma}\tilde{R}$$
$$R:=\tilde{R}^{-1}\ ,$$
we find that also
\begin{equation}
(\Gamma,R,w)\mapsto \A_1(\Gamma)\A_2(R)w\
\end{equation}
is a 24-fold covering of $\F$, which completes the proof of (ii) and of Theorem \ref{coordinates}.

\section{Kinematics}
\subsection{Elastic potential energy}

Let us study the form of the elastic potential energy and of the
potential energy of self-gravitation in the coordinates just introduced. With an
abuse of terminology, we will call the sum of these two potential
energies simply ``elastic potential energy'' and we will denote it by
$V_e$; furthermore, we will refer to the corresponding forces as to
the ``elastic forces'', leaving understood that they include also the forces
related to self-gravitation.

\begin{hyp}
The identical deformation $v_0(\bx)=\bx$ is a minimum of the
elastic (and self-gravitational) potential energy. 
\end{hyp}

\begin{remark}
This Assumption means that we have chosen the material space $\Omega$ as representing an abstract realization of the equilibrium configuration of the compressed elastic sphere under the effect of self-gravitation (and not the totally undeformed elastic body).
\end{remark}

In the equilibrium state, because of the rotational invariance, all three
principal moments of inertia are equal to the same constant $I_0$. For
simplicity, we use the differences between the $I_j$'s and $I_0$ as
configuration variables instead of the $I_j$'s themselves, so we define
\begin{equation}
J_i:=I_i-I_0\ , \qquad \qquad (i=1,2,3)\ ,
\end{equation}
and we assume (without loss of generality) that 
$$
z_j(v_0)=0\quad \forall j\ .
$$

\begin{remark}
Due to the $\mathcal{A}_1$- and $\mathcal{A}_2$-invariance, the
elastic potential energy does not depend on $\Gamma$ and $R$.
\end{remark}

We also assume that the minimum is nondegenerate and that the body is
very rigid. Summarizing we make the following Assumption:

\begin{hyp}
The elastic potential energy has the form
\begin{equation}
\label{pot.hip}
V_e(J,z)=\frac{1}{\varepsilon}V_0(J,z)\equiv
\frac{1}{\varepsilon}\left[ Q(J,z) +V_3(J,z)\right]\ ,
\end{equation}
where $\varepsilon$ is a small parameter, $Q$ a nondegenerate
quadratic form and $V_3$ has a zero of order three at the origin.
\end{hyp}

We want to study more in detail the form of the elastic potential near
the equilibrium, but we have to cope with the fact that our
coordinates are singular at the equilibrium configuration $v_0(\bx)=\bx$. 

\begin{lemma}
The elastic potential energy has the form
\begin{eqnarray} \label{elasticpot}
Q(J,z) & = & \frac{A}{2}({J_1}^2+{J_2}^2+{J_3}^2)+B(J_1 J_2+J_1
J_3+J_2 J_3)+
\\
\nonumber & & +\sum_{j=1}^n C_j z_j
(J_1+J_2+J_3)+\frac{1}{2}\sum_{j,k=1}^n D_{jk} z_j z_k \ ,
\end{eqnarray}
where the constants $A,B,C_j,D_{jk}$ are such that the quadratic part
$Q(J,z)$ is a positive definite quadratic form in the variables
$(J,z)$. In particular, this implies $A>B$.
\end{lemma}

\begin{remark}
By Remark \ref{I.sing}, such an expression can be used to compute the Lagrange equations only outside $\C_{=}$.
\end{remark}

\begin{proof}
Any $v\in\mathcal{F}$ can be represented as
$\mathcal{A}_1(\Gamma)\mathcal{A}_2(R)w$, for some $\Gamma,R\in SO(3)$ and $w\in\Sc\cap\C_{\neq}$. Moreover, due to the
group action invariance, the potential energy associated to the
configuration $v$ must be the same as the potential energy
associated to the configuration $w$. Therefore the potential energy is a function of the variables $J,z$ only, it can be computed
working in $\Sc$ and then the obtained form holds on the whole of
$\F$.

The rotational invariance implies that the expression of the elastic potential energy must be symmetric with respect to permutations of
the indices $i=1,2,3$, and the expression of $Q(J,z)$ in
(\ref{elasticpot}) is the most general expression of a quadratic form
with such a property.
\end{proof}

\subsection{Planar restriction}

We are going to study the dynamics of the satellite in the special case when the spin axis of the body is orthogonal to the plane of the orbit and coincides with one of the principal axes of inertia of the body. For this reason, in the rest of the paper we will restrict to ``planar deformations'' and ``planar rotations''. Precisely, we make the following assumptions.

\begin{hyp}[Planar deformations] \label{plandef}
The configuration is such that $\be_3$ is an eigenvector of $I$. We label the principal axes of inertia so that $\bu_3=\be_3$.
\end{hyp}

In other words, we are assuming the matrix of inertia of the satellite
to be of the form
\begin{equation}
I(v)=\left[ \begin{array}{ccc}
I_{11}(v) & I_{12}(v) & 0 \\
I_{12}(v) & I_{22}(v) & 0 \\
0 & 0 & I_{33}(v) \end{array} \right]\ ,
\end{equation}
so that $I_3(v)\equiv I_{33}(v)$.

\begin{hyp}[Planar rotations] \label{planrot}
$\Gamma$ is a rotation about the $\be_3$-axis, i.e. it has the form
\begin{equation}
\Gamma=\Gamma(\alpha):=\left[ \begin{array}{ccc}
\cos\alpha & -\sin\alpha & 0 \\
\sin\alpha & \cos\alpha & 0 \\
0 & 0 & 1 \end{array} \right]\ .
\end{equation}
\end{hyp}

As a consequence of these two assumptions, $R$ is also a rotation about the $\be_3$-axis, i.e. there exists an angle $\beta$ such that
\begin{equation}
R=R(\beta):=\left[ \begin{array}{ccc}
\cos\beta & -\sin\beta & 0 \\
\sin\beta & \cos\beta & 0 \\
0 & 0 & 1 \end{array} \right]\ .
\end{equation}

\begin{remark}
The assumptions \ref{plandef} and \ref{planrot}, together with Theorem
\ref{coordinates}, imply that
$(\alpha,\beta,I_1,I_2,I_3,z_1,z_2,\ldots,z_n)$ are good Lagrangian
coordinates for the body configuration. Actually, by following the
proof of Theorem \ref{coordinates} one can show that such coordinates
form a 4-fold covering of the configuration space restricted to planar
configurations.  

The fact that dynamics remains confined for all times within the set
$\mathcal{F}$ will be guaranteed by the local stability result proved
in the following sections.
\end{remark}

\subsection{Kinetic energy}

By K\"onig's second Theorem, the kinetic energy $T$ can be written as
the sum of two terms: the former is the kinetic energy of the center
of mass
\begin{equation} \label{eq:Tcm}
T_{cm}=\frac{m}{2}({\dot R}^2+R^2 {\dot\psi}^2)\ ,
\end{equation}
where $R,\psi$ are the polar coordinates of the center of mass $\bR$, and the latter is the kinetic energy of the satellite with respect to its center of mass
\begin{equation} \label{eq:Tr}
T_r:=T_r(\dot\alpha,\dot\beta,\dot{J_1},\dot{J_2},\dot{J_3},\dot{z};J_1,J_2,J_3,z)\ .
\end{equation}

\begin{remark}
$T_r$ is independent of $\alpha$ and $\beta$ due to the rotational invariance of the satellite.
\end{remark}

We will use the notation
\begin{equation}
T_r:=\frac{1}{2}\sum_{i,k=1}^{5+n} a_{ik}(J,z) {\dot q}_i {\dot q}_k\ ,
\end{equation}
where $q=(\alpha,\beta,J_1,J_2,J_3,z)$. Observe that the coefficients $a_{ik}(J,z)$ are such that the quadratic form is positive definite on $\F$.

\begin{lemma} \label{kin.lemma}
The coefficients $a_{ik}(J,z)$ satisfy:
\begin{description}
	\item[(i)] \begin{equation}
	a_{11}(J,z)=I_3=I_0+J_3
	\end{equation}
	\item[(ii)] \begin{equation}\label{a12}
	\lim_{(J,z)\rightarrow 0}{a_{12}(J,z)}=0\ .
	\end{equation}
\end{description}
\end{lemma}

\begin{proof}
For $w\in \Sc$, we set
\begin{equation}
w_{def}(\bx):=w(\bx)-v_0(\bx)=w(\bx)-\bx\ .
\end{equation}
and
\begin{equation}
u=\mathcal{A}_2(R(\beta))w_{def}\ .
\end{equation}
We remark that $\mathcal{A}_2(R(\beta))v_0=v_0$.

Now, let us evaluate the kinetic energy $T_r$. For $v\in\F$ we have
\begin{equation}
v=\Gamma(\alpha)(\bx+u(\bx))\ .
\end{equation}
Taking the derivative with respect to time, we get
\begin{equation}
\dot{v}(\bx)=\frac{d \Gamma(\alpha)}{d t}(\bx+u(\bx))+\Gamma(\alpha)\dot{u}(\bx)\ .
\end{equation}
Therefore,
\begin{eqnarray} \label{kinenergy}
\nonumber T_r & = & \frac{1}{2}\int_{\Omega}{[\dot{v}(\bx)]^2 \rho(\bx)d^3}=\frac{1}{2}\int_{\Omega}{[\Gamma(-\alpha)\dot{v}(\bx)]^2 \rho(\bx)d^3 \bx}=\\
\nonumber & = & \frac{1}{2}\int_{\Omega}{\left[\Gamma(-\alpha)\frac{d\Gamma(\alpha)}{d t}(\bx+u(\bx))+\dot{u}(\bx)\right]^2 \rho(\bx)d^3 \bx}=\\
\nonumber & = & \frac{1}{2}\int_{\Omega}{\left\{\omega\times\left[\bx+u(\bx)\right]\right\}^2 \rho(\bx)d^3 \bx} +\\
& & + \int_{\Omega}{\langle\omega\times\left[\bx+u(\bx)\right],\dot{u}(\bx)\rangle \rho(\bx)d^3 \bx} + \frac{1}{2}\int_{\Omega}{\left[\dot{u}(\bx)\right]^2 \rho(\bx)d^3 \bx},
\end{eqnarray}
where $\omega$ is the angular velocity of the satellite, defined by  $\omega\times(\cdot)=[\Gamma(-\alpha)][\frac{d}{d t}\Gamma(\alpha)](\cdot)$. Under our assumptions, $\omega=\dot{\alpha}\bu_3$.

As the vector field $u(\bx)$ is independent of $\alpha$, we observe that $T_r$ is the sum of three integrals, the first of which gives the term in ${\dot\alpha}^2$, while the third one is a quadratic form in $(\dot\beta,\dot{J_1},\dot{J_2},\dot{J_3},\dot z)$ and the second one gives mixed terms in $\dot\alpha$ and in the other velocities.

Therefore, one gets
\begin{equation}
a_{11}(J,z)=\int_{\Omega}{\left\{\bu_3 \times\left[\bx+u(\bx)\right]\right\}^2 \rho(\bx)d^3 \bx}\ ,
\end{equation}
and it can easily be seen that this expression equals the moment of inertia related to the vertical axis, thus proving (i).

For the second part of the Lemma, we have to study the coefficient of the term $\dot{\alpha}\dot{\beta}$. Observe that such a term arises from the integral $$\int_{\Omega}{\langle\omega\times\left[\bx+u(\bx)\right],\dot{u}(\bx)\rangle \rho(\bx)d^3 \bx}$$ in (\ref{kinenergy}). Here, the $\dot{\alpha}$ factor comes from the angular velocity $\omega$, while the $\dot{\beta}$ factor is hidden in $\dot{u}(\bx)$. Using
\begin{equation}
u(\bx)=R(\beta)w_{def}[R(-\beta)\bx]\ ,
\end{equation}
we get
\begin{eqnarray} \label{eq:dotu}
\nonumber \dot{u}(\bx) & = & R(\beta)\left\{\dot{w}_{def}\left[R(-\beta)\bx\right]\right\}+\dot{\beta}\frac{\partial R(\beta)}{\partial\beta}w_{def}\left[R(-\beta)\bx\right]+\\
& & +\dot{\beta}R(\beta)\nabla w_{def}\left[R(-\beta)\bx\right]\cdot \frac{\partial R(-\beta)}{\partial\beta}\bx\ .
\end{eqnarray}
Here, we notice that the first of the three addenda is independent of $\dot{\beta}$, so we have
\begin{eqnarray}
\nonumber a_{12}(J,z) & = & \frac{1}{2}\int_{\Omega}{\langle\bu_3\times\left\{\bx+R(\beta)w_{def}[R(-\beta)\bx]\right\}},\frac{\partial R(\beta)}{\partial\beta}w_{def}\left[R(-\beta)\bx\right]+\\
& & \phantom{\int_{\Omega}}+R(\beta)\nabla w_{def}\left[R(-\beta)\bx\right]\cdot \frac{\partial R(-\beta)}{\partial\beta}\bx\rangle \rho(\bx)d^3 \bx\ ,
\end{eqnarray}
which goes to zero when $w_{def} \rightarrow 0$, i.e. when $(J,z) \rightarrow 0$; this completes the proof of (ii).
\end{proof}

\subsection{Gravitational potential energy}

Now we evaluate the gravitational potential energy of the satellite subject to the gravitational field
of a pointlike center (planet) having mass $M$.

To start with, we fix some notation: $(R,\psi)$ are the polar coordinates of the center of mass of the satellite in the plane of the orbit.
We denote with $\gamma$ the angle between the principal axis $\bu_1$ and the line joining the planet to the center of mass of the satellite. Such a line is usually referred to as the \emph{line of centres}. Observe that $\gamma=\alpha+\beta-\psi$. Furthermore, we set $\chi:=\alpha-\psi$, i.e. $\chi$ is the angle of rigid rotation of the satellite, measured with respect to the line of centres.

\begin{proposition}
\label{p.1}
In the quadrupole approximation the gravitational potential energy of the
body in the field generated by the mass $M$ is given by
\begin{equation}
\label{e.2}
V_g(\bR,J_1,J_2,J_3,\gamma):=-\frac{GMm}{R}+\frac{GM}{R^3}[-J_1+2
  J_2-J_3+3(J_1-J_2)\cos^2{\gamma}]
\end{equation}
\end{proposition}

\begin{proof}
See Appendix \ref{gravitation}. A proof of this result can also be found in the book \cite{bertotti}.
\end{proof}

\section{Conservative dynamics}

After having described our model of satellite, we are ready to study the dynamics of our system.
At first, we neglect the dissipative effects and study the dynamics of the corresponding conservative system.

The Lagrangian
\begin{equation}
\lag=T-V
\end{equation}
of the system is the difference between the kinetic energy $T$ and the total potential energy.
Therefore, collecting the above results one has:
\begin{eqnarray}\label{eq:lagrangian}
\lag & = & \frac{m}{2}\left(\dot{R}^2+R^2 \dot{\psi}^2\right)+T_r(\dot\chi+\dot\psi,\dot{J},\dot\beta,\dot{z};J,z)+\\
\nonumber & & +\frac{G M m}{R}+\frac{G M}{R^3}\left[J_1-2 J_2+J_3+3(J_2-J_1)\cos^2(\chi+\beta)\right]-V_e(J,z; \varepsilon)\ .
\end{eqnarray}
Now, we observe that the Lagrangian does not depend on the cyclic
coordinate $\psi$, so the total angular momentum
\begin{equation} \label{eq:p}
p:=\frac{\partial\lag}{\partial\dot{\psi}}=m R^2 \dot{\psi}+(J_3+I_0)(\dot{\chi}+\dot{\psi})+2\sum_{k=2}^{5+n} a_{1k}(J,z){\dot q}_k\ ,
\end{equation}
is a constant of motion. We can invert relation (\ref{eq:p}), to get the expression of $\dot{\psi}$ as a function of the other variables:
\begin{equation}
\dot{\psi}=\frac{p-(J_3+I_0)\dot{\chi}-2\sum_{k=2}^{5+n} a_{1k}(J,z){\dot q}_k}{m R^2+I_0+J_3}\ .
\end{equation}
Then, we can drop one degree of freedom and study the reduced Lagrangian
\begin{equation}
\lagr=\lag-\dot{\psi}\frac{\partial\lag}{\partial\dot{\psi}}\ ,
\end{equation}
where $\dot{\psi}$ must be thought of as a function of the other
Lagrangian coordinates and velocities. After some calculations, we get
\begin{equation}
\lagr=T_2+T_1-\Veff\ ,
\end{equation}
where
\begin{eqnarray}
T_2 & = & \frac{m}{2}\dot{R}^2+T_r(\dot\chi,\dot{J},\dot\beta,\dot{z};J,z)-\frac{\left[(J_3+I_0)\dot{\chi}+2\sum_{k=2}^{5+n} a_{1k}(J,z){\dot q}_k\right]^2}{2(m R^2+I_0+J_3)}\\
T_1 & = & \frac{p\left[(J_3+I_0)\dot{\chi}+2\sum_{k=2}^{5+n} a_{1k}(J,z){\dot q}_k\right]}{m R^2+I_0+J_3}\\
\Veff & = & \frac{p^2}{2(m R^2+I_0+J_3)}+\\
\nonumber & & -\frac{G M m}{R}-\frac{G M}{R^3}\left[J_1-2 J_2+J_3+3(J_2-J_1)\cos^2{\gamma}\right]+V_e(J,z; \varepsilon)\ .
\end{eqnarray}

Such a system has the conserved quantity
\begin{equation}
E:=T_2+\Veff=\sum_{k=1}^{5+n}\dot{y}_k\frac{\partial\lagr}{\partial\dot{y}_k}-\lagr\ ,
\end{equation}
where $$y:=(R,\chi,\beta,J_1,J_2,J_3,z_1,z_2,\ldots)$$ and {\it the strict minima of $\Veff$ are Lyapunov-stable equilibria of the system}.

Let $R_0$ be a nondegenerate minimum of the function 
\begin{equation}
\label{vG}
V_{G0}(R):=-\frac{GMm}{R}+\frac{p^2}{2(mR^2+I_0)}\ .
\end{equation}
Then we have the following
\begin{lemma}
For any $\varepsilon$ small enough, there exist $\bar R, \bar J,\bar
z$, s.t. 
\begin{itemize}
\item[(1)] 
the manifold
$$\mathcal{M}:=\left\{(\bar{R},\chi,\beta,\bar{J}_1,
\bar{J}_2,\bar{J}_3,\bar{z})|\chi+\beta=0\right\}\ ,$$
is composed by critical points of $\Veff$. 
\item[(2)]
$\M$ is a
minimum of $\Veff$ which is nondegenerate in the transversal direction.
\item[(3)] One has $(\bar{J}_1,
\bar{J}_2,\bar{J}_3,\bar{z})=O(\varepsilon)$ and $|\bar
R-R_0|=O(\varepsilon)$. 
\item[(4)] Finally $\bar{J}_1<\bar{J}_2<\bar{J}_3$.
\end{itemize}
\end{lemma}

\begin{remark}
Point (iv) guarantees that $\mathcal{M}\subset\C_{\neq}$. If $\varepsilon$ is sufficiently small, then we have $\mathcal{M}\subset\mathcal{F}$.
\end{remark}

\begin{remark}
$\mathcal{M}$ is the manifold corresponding to 1:1 spin orbit resonance.
\end{remark}

\begin{proof}  We look for a minimum of $\tilde V$ in the domain
  $J_1\leq J_2$ and $|J|\leq C\varepsilon$ for some fixed $C$.

First remark that, as a function of $\gamma=\chi+\beta$, $\tilde V$ has a minimum
at $\gamma$=0 (strict if $J_1< J_2$). Consider now $\tilde
V\big|_{\gamma=0}$; as a function of $R$ it has a nondegenerate minimum
at some point $R=R(J,z)$ fulfilling 
$$
\left| R(J,z)-R_0\right|\leq C\varepsilon\ .
$$
Consider now the restriction $\bar V=\bar V(J,z)$ of $\tilde V$ to the
manifold $\gamma=0,$\ $R=R(J,z)$;
since
\begin{equation}
\bar V(J,z)= 
\frac{1}{\varepsilon}\left[Q(J,z)+V_3(J,z)\right]+O(1)\ ,
\end{equation}
such a function has a nondegenerate minimum close to zero.

Then (1), (2) and (3) follow provided one shows that
$\bar{J}_1<\bar{J}_2$. We are now going to prove (4) which in
particular implies the thesis.

To this end, observe that at the critical point one has
\begin{equation} \label{eq:pos1}
0=\frac{\partial \tilde V}{\partial J_1}=\frac{2 G
  M}{\bar{R}^3}+\frac{A}{\varepsilon}\bar{J}_1+\frac{B}{\varepsilon}(\bar{J}_2+\bar{J}_3)+\frac{1}{\varepsilon}\sum_{j=1}^{n}
C_j \bar{z}_j+O(\varepsilon)
\end{equation}
\begin{equation} \label{eq:pos2}
0=\frac{\partial \tilde V}{\partial J_2}=-\frac{G
  M}{\bar{R}^3}+\frac{A}{\varepsilon}\bar{J}_2+\frac{B}{\varepsilon}(\bar{J}_1+\bar{J}_3)+\frac{1}{\varepsilon}\sum_{j=1}^{n}
C_j \bar{z}_j+
O(\varepsilon)\ .
\end{equation}
\begin{eqnarray} \label{eq:pos3}
\nonumber 0 & = & \frac{\partial \tilde V}{\partial
  J_3}=-\frac{p^2}{2(m\bar{R}^2+I_0+\bar{J}_3)^2}-\frac{G
  M}{\bar{R}^3}+\\
  & & +\frac{A}{\varepsilon}\bar{J}_3+\frac{B}{\varepsilon}(\bar{J}_1+\bar{J}_2)+\frac{1}{\varepsilon}\sum_{j=1}^{n}
C_j \bar{z}_j+O(\varepsilon)\ .
\end{eqnarray}
Subtracting (\ref{eq:pos2}) from
(\ref{eq:pos1}), we obtain
\begin{equation}
\frac{3 G
  M}{\bar{R}^3}+\frac{A-B}{\varepsilon}(\bar{J}_1-\bar{J_2})+O(\varepsilon)=0\ .
\end{equation}
The positive definiteness of the quadratic form $Q$ implies
$A-B>0$. Therefore, if $\varepsilon$ is sufficiently small, we have
$\bar{J}_1<\bar{J}_2$.
Subtracting (\ref{eq:pos3}) from (\ref{eq:pos2}), we get
\begin{equation}
\frac{p^2}{2(m\bar{R}^2+I_0+\bar{J}_3)^2}+\frac{A-B}{\varepsilon}(\bar{J}_2-\bar{J_3})+O(\varepsilon)=0\ .
\end{equation}
Hence, for $\varepsilon$ sufficiently small, we have
$\bar{J}_2<\bar{J}_3$.
\end{proof}

\begin{coro}
The critical manifold $\mathcal{M}$ is an orbitally stable equilibrium
for the Lagrangian system of equations
\begin{equation} \label{eq:lag}
\frac{d}{dt}\frac{\partial\lagr}{\partial\dot{y}_k}=
\frac{\partial\lagr}{\partial
  y_k}\ . \qquad \qquad (k=1,2,\ldots)
\end{equation}
\end{coro}

\section{Dissipative dynamics}
\label{dissi}

In the previous section, we have studied the dynamical properties of the conservative system. However, we are interested in considering the effects on the dynamics caused by friction within the satellite.

To this end, we modify the Euler-Lagrange equations \eqref{eq:lag} by adding to the r.h.s. the terms $-f_k(\dot y,y)$. Namely we study the equations
\begin{equation} \label{eq:dissip}
\frac{d}{dt}\frac{\partial\lagr}{\partial\dot{y}_k}-
\frac{\partial\lagr}{\partial
  y_k}=-f_k(\dot y,y)\ , \qquad (k=1,2,\ldots,n+6)
\end{equation}
where we assume that the functions $f_k$ are continuous in the arguments $(\dot y,y)$.

Now observe that, in the system of equations \eqref{eq:dissip}, the Lie derivative of the energy is given by
$$\frac{dE}{dt}=\frac{d}{dt}\sum_{k=1}^{n+6}\left(\frac{\partial\lagr}{\partial\dot y_k}\dot y_k-\lagr\right)=\sum_{k=1}^{n+6}\left[\left(\frac{d}{dt}\frac{\partial\lagr}{\partial\dot{y}_k}-\frac{\partial\lagr}{\partial y_k}\right)\dot y_k\right]=-\sum_{k=1}^{n+6}\dot{y}_k f_k(\dot y,y)\ .$$

The property that the $f_k$-terms represent a dissipative force is therefore summarized by the following
\begin{hyp}
The functions $f_k$ satisfy
\begin{equation}\label{dissipation.function}
\sum_{k=1}^{n+6}\dot{y}_k f_k(\dot y,y)\geq 0\ .
\end{equation}
\end{hyp}

Moreover, we assume that \emph{any} deformation of the body dissipates some energy. We define $y^e:=(\beta,J,z)$ to be the vector of the variables describing the body deformation. Then we state the following assumption.

\begin{hyp}\label{y.e}
The non-dissipation condition
\begin{equation}
\sum_{k=1}^{n+6}\dot{y}_k f_k(\dot y,y)=0
\end{equation}
is satisfied if and only if
$$\dot\beta=\dot J=\dot z=0\ ,$$
i.e. if $\dot{y}^e=0$.
\end{hyp}

\begin{remark}
Assumption \ref{y.e} implies that the terms $f_1$ and $f_2$ corresponding to the variables $R$ and $\chi$ are identically zero. Moreover, by continuity of the $f_k$'s, this also implies $f_k=0$ whenever $\dot{y}^e=0$.
\end{remark}

Now we can state our main result:
\begin{teo} \label{asympt.stab}
If $\varepsilon$ is sufficiently small, the manifold $\mathcal{M}$,
defined in the previous section, is an asymptotically stable
equilibrium for the dynamical system of equations \eqref{eq:dissip}.
\end{teo}

In order to prove this Theorem, we will use a common tool in the study
of dynamical systems, known as \emph{LaSalle's invariance principle} (see \cite{lasalle}, \S 2.6),
which allows one to prove results of asymptotical stability in presence of
a Lyapunov function $E$ satisfying a nonstrict inequality of the type
$\dot{E}\leq 0$. To state this principle, we first recall the
definition of an invariant set.
\begin{defi}
A subset $M$ of the phase space is called (positively) invariant if
all solutions starting in $M$ remain in $M$ for all future times.
\end{defi}
We now state a version of LaSalle's Theorem.
\begin{teo}[LaSalle's invariance principle]
Suppose that $E$ is a real-valued smooth function defined on the phase
space, satisfying $\dot{E}(y,\dot{y})\leq 0$ for all
$(y,\dot{y})$. Let $M$ be the largest invariant set contained in
$\left\{(y,\dot{y})|\dot{E}(y,\dot{y})=0\right\}$. Then every solution
that remains within a compact subset of the phase space for $t\geq 0$ approaches $M$ as $t\rightarrow +\infty$.
\end{teo}

\begin{proof}[Proof of Theorem \ref{asympt.stab}]
Let $\mathcal{ND}:=\left\{(y,\dot{y})|\dot{y}^e=0\right\}$ be the subset
of phase space where there is no energy dissipation.  Then, due to
LaSalle's invariance principle, any solution such that
$(y(0),\dot{y}(0))$ belongs to a sufficiently small neighborhood of
$(\mathcal{M},0)$ (notice that such a solution will stay bounded for all $t\geq 0$
due to the Lyapunov stability of $\mathcal{M}$ which has been proved
in the previous section) will get arbitrarily close to the largest
invariant subset of $\mathcal{ND}$, for $t\rightarrow
+\infty$. Therefore, the only thing we have to check is that the set $\mathcal{ND}$
contains no orbit, apart from the points of the manifold
$\mathcal{M}$. To check this, observe that, if such an orbit existed,
it would satisfy equations (\ref{eq:dissip}). In particular, the orbit satisfies
\begin{equation}
\frac{d}{dt}\frac{\partial\lagr}{\partial\dot\chi}-\frac{\partial\lagr}{\partial\chi}=0
\end{equation}
and
\begin{equation}
\frac{d}{dt}\frac{\partial\lagr}{\partial\dot\beta}-\frac{\partial\lagr}{\partial\beta}=-\frac{\partial F}{\partial\dot\beta}\ .
\end{equation}
When restricting to $\mathcal{ND}$, these two equations become, respectively,
\begin{eqnarray} \label{eq.diss.1}
-\frac{(I_0+J_3)^2\ddot\chi}{m R^2+I_0+J_3}+\frac{2m R(I_0+J_3)^2\dot\chi\dot R}{(m R^2+I_0+J_3)^2}+ & &\\
\nonumber +(I_0+J_3)\ddot\chi-\frac{2p m R(I_0+J_3)\dot R}{(m R^2+I_0+J_3)^2} & = & -\frac{\partial\Veff}{\partial\gamma}(y)
\end{eqnarray}
and
\begin{eqnarray} \label{eq.diss.2}
-\frac{a_{12}(J,z)(I_0+J_3)\ddot\chi}{m R^2+I_0+J_3}+\frac{2m R a_{12}(J,z)(I_0+J_3)\dot\chi\dot R}{(m R^2+I_0+J_3)^2}+ & &\\
\nonumber +a_{12}(J,z)\ddot\chi-\frac{2p m R a_{12}(J,z)\dot R}{(m R^2+I_0+J_3)^2} & = & -\frac{\partial\Veff}{\partial\gamma}(y)\ ,
\end{eqnarray}
where we took into account that the $f_k$'s vanish on $\mathcal{ND}$. 

Multipliying (\ref{eq.diss.1}) by 
$\frac{a_{12}(J,z)}{I_0+J_3}$ and subtracting \eqref{eq.diss.2} we get
$$
\left(1-\frac{a_{12}}{I_0+J_3}\right)\frac{\partial \tilde
  V}{\partial \gamma}= 0\ ,
$$ which, by \eqref{a12}, implies $\frac{\partial \tilde V}{\partial
  \gamma}= 0$, and therefore $\chi+\beta=0$. Then, we have
$\dot\chi=-\dot\beta=0$, since $\dot\beta=0$ on $\mathcal{ND}$. Now ,
substituting $\dot\chi=\ddot\chi=0$ into equations
(\ref{eq.diss.1}) and (\ref{eq.diss.2}), we find $\dot R=0$. Finally,
we observe that now we have
\begin{equation}
\dot\chi=\dot\beta=\dot{R}=\dot{J}=\dot{z}=0\ ,
\end{equation}
which is true on the equilibrium manifold $\mathcal{M}$ only. 

We have thus proved that the only orbits contained in $\mathcal{ND}$
are the points of the 1:1 spin-orbit resonance manifold $\mathcal{M}$,
which implies the asymptotic stability of $\mathcal{M}$.
\end{proof}

This concludes the proof of the asymptotic stability of the synchronous resonance for the system with dissipation.

\appendix{Gravitational potential energy}\label{gravitation}

\textit{Proof of Proposition \ref{p.1}}: Define $\tilde{\rho}:\zeta(\Omega) \longrightarrow \mathbb{R}$
as $$\tilde{\rho}(\xi):=\frac{\rho(\zeta^{-1}(\xi))}{\left|
  \det\frac{\partial\zeta}{\partial\bx}(\zeta^{-1}(\xi)) \right|}\ ,$$
i.e. $\tilde{\rho}(\xi)$ is the density of the satellite at the point
$\xi=\zeta(\bx)$.  Let $(x_1, x_2, x_3)$ be the Cartesian coordinates
referred to the system with origin $\bR$ and axes $\bu_1 \bu_2
\bu_3$. Then we introduce the spherical coordinates
$(r,\vartheta,\phi)$ of the generic point $\bP$ in the satellite,
defined by:
\begin{eqnarray} \label{eq:spherical}
x_1 & = & r \cos{\vartheta}\cos{\phi}\\
x_2 & = & r \sin{\vartheta}\cos{\phi}\\
x_3 & = & r \sin{\phi}\ .
\end{eqnarray}
In this frame, the products of inertia $I_{ij}$ vanish, i.e.
\begin{eqnarray} \label{in.2}
\int_{\zeta(\Omega)}{\tilde{\rho}(\xi) r^2 \cos^2{\phi}\cos{\vartheta}\sin{\vartheta}d^3\xi} & = & 0\\
\int_{\zeta(\Omega)}{\tilde{\rho}(\xi) r^2 \cos{\phi}\sin{\phi}\cos{\vartheta}d^3\xi} & = & 0\\
\int_{\zeta(\Omega)}{\tilde{\rho}(\xi) r^2 \cos{\phi}\sin{\phi}\sin{\vartheta}d^3\xi} & = & 0\ ,
\end{eqnarray}
and the principal moments of inertia are given by
\begin{eqnarray} \label{in.3}
I_1 & = & I_0+J_1=K_2+K_3\\
I_2 & = & I_0+J_2=K_1+K_3\\
I_3 & = & I_0+J_3=K_1+K_2\ ,
\end{eqnarray}
where
\begin{eqnarray} \label{in.4}
K_1 & = & \frac{1}{2}\int_{\zeta(\Omega)}{\tilde{\rho}(\xi) r^2 \cos^2{\phi}\cos^2{\vartheta}d^3\xi}\\
K_2 & = & \frac{1}{2}\int_{\zeta(\Omega)}{\tilde{\rho}(\xi) r^2 \cos^2{\phi}\sin^2{\vartheta}d^3\xi}\\
K_3 & = & \frac{1}{2}\int_{\zeta(\Omega)}{\tilde{\rho}(\xi) r^2 \sin^2{\phi}d^3\xi}\ .
\end{eqnarray}
The gravitational potential energy $V_g$ is:
\begin{equation}
V_g=-\int_{\zeta(\Omega)}{\frac{G M \tilde{\rho}(\xi)}{|\xi|}d^3\xi}=-\int_{\zeta(\Omega)}{\frac{G M \tilde{\rho}(\xi)}{\sqrt{R^2+r^2-2 R r \cos{\eta}}}d^3\xi}\ ,
\end{equation}
where $\eta$ is the angle between the line of centres and $\bR \bP$. Notice that the relation
\begin{equation} \label{eq:etaphithetagamma}
\cos{\eta}=\cos{\phi}\cos{(\vartheta+\gamma)}
\end{equation}
holds. Let us recall now how the multipole expansion arises. We have
\begin{equation}
\frac{1}{|\xi|}=\frac{1}{\sqrt{R^2+r^2-2 R r \cos{\eta}}}=\frac{1}{R}\frac{1}{\sqrt{1+\left(\frac{r}{R}\right)^2-2\left(\frac{r}{R}\right)\cos{\eta}}}\ .
\end{equation}
In terms of the Legendre polynomials $P_n(z)$, one has
\begin{equation}
\frac{1}{\sqrt{1+x^2-2 x z}}=\sum_{n\geq 0}{x^n P_n(z)}\ .
\end{equation}
In particular, we recall that $$P_0(z)=1$$ $$P_1(z)=z$$ $$P_2(z)=\frac{3 z^2 -1}{2}\ .$$
Taking the quadrupole approximation means to cut the sum at $n=2$.
We get
\begin{equation}
\frac{1}{|\xi|}=\frac{1}{R}\sum_{n\geq 0}{\left(\frac{r}{R}\right)^n P_n(\cos{\eta})}\simeq \frac{1}{R}\left[1+\frac{r}{R}\cos{\eta}+\left(\frac{r}{R}\right)^2\frac{3\cos^2{\eta}-1}{2}\right]\ ,
\end{equation}
so the potential energy becomes
\begin{equation}
V_g=-\int_{\zeta(\Omega)}{\frac{G M \tilde{\rho}(\xi)}{R}\left[1+\frac{r}{R}\cos{\eta}+\left(\frac{r}{R}\right)^2\frac{3\cos^2{\eta}-1}{2}\right]d^3\xi}\ .
\end{equation}
Here, the first term equals $-\frac{G M m}{R}$; the second one vanishes because $\bR$ is the center of mass of the satellite; the third term, namely $$V_t:=-\int_{\zeta(\Omega)}{\frac{G M \tilde{\rho}(\xi)(3\cos^2{\eta}-1)r^2}{2 R^3}d^3\xi}\ ,$$
gives what we call the ``tidal'' potential energy.
A brief manipulation shows that
\begin{eqnarray}
\nonumber V_t & = & -\frac{G M}{R^3}\int_{\zeta(\Omega)}{\frac{\tilde{\rho}(\xi)(3\cos^2{\eta}-1) r^2}{2}d^3\xi}=\\
\nonumber & = & -\frac{G M}{2 R^3}\int_{\zeta(\Omega)}{\tilde{\rho}(\xi) r^2[3\cos^2{\phi}(\cos{\vartheta}\cos{\gamma}-\sin{\vartheta}\sin{\gamma})^2-1]d^3\xi}=\\
\nonumber & = & -\frac{3 G M}{R^3}(K_1\cos^2{\gamma}+K_2\sin^2{\gamma})+\frac{G M}{2 R^3}\int_{\zeta(\Omega)}{\tilde{\rho}(\xi) r^2 d^3\xi}=\\
\nonumber & = & -\frac{3 G M}{R^3}(K_1\cos^2{\gamma}+K_2\sin^2{\gamma})+\\
\nonumber & & + \frac{G M}{2 R^3}\int_{\zeta(\Omega)}{\tilde{\rho}(\xi) r^2 [\sin^2{\phi}+\cos^2{\phi}(\sin^2{\vartheta}+\cos^2{\vartheta})] d^3\xi}=\\
\nonumber & = & -\frac{3 G M}{R^3}(K_1\cos^2{\gamma}+K_2\sin^2{\gamma})+\frac{G M}{R^3}(K_1+K_2+K_3)=\\
& = & \frac{G M}{R^3}[-J_1+2 J_2-J_3+3(J_1-J_2)\cos^2{\gamma}]\ .
\end{eqnarray}

\vskip15pt

{\it  E-mail address:} {\tt dario.bambusi@unimi.it}

\vskip8pt

\noindent Dipartimento di Matematica ``F. Enriques'', Universit\`a
degli Studi di Milano, Via Saldini 50, 20133 Milano, Italy.

\vskip15pt

{\it  E-mail address:} {\tt emanuele.haus@univ-nantes.fr}

\vskip8pt

\noindent Laboratoire de Mathématiques Jean Leray, Université
de Nantes, \\2 rue de la Houssinière, 44322 Nantes, France.

\end{document}